\documentclass{article}

\usepackage{PRIMEarxiv}

\usepackage[utf8]{inputenc} 
\usepackage[T1]{fontenc}    
\usepackage{hyperref}       
\usepackage{url}            
\usepackage{booktabs}       
\usepackage{amsfonts}       
\usepackage{amssymb}        
\usepackage{amsthm}         
\usepackage{amsmath}
\usepackage{extarrows}      
\usepackage{nicefrac}       
\usepackage{microtype}      
\usepackage{lipsum}
\usepackage{fancyhdr}       
\usepackage{graphicx}       
\usepackage{threeparttable} 

\usepackage{setspace}       

\graphicspath{{media/}}     

\newtheorem{proposition}{Proposition}  

\title{Conditional Estimates of Diffusion Processes for Evaluating the Positive Feedback Trading
\date{May, 2019}
}

\author{
  Aihua Li \\
  Department of Statistics and Management \\
  Shanghai University of Finance and Economics \\
  \texttt{aihua.li@duke.edu}
}

\begin{document}
\maketitle

\begin{abstract}
  Positive feedback trading, which buys when prices rise and sells when prices fall, has long been criticized for being destabilizing as it moves prices away from the fundamentals.
  Motivated by the relationship between positive feedback trading and investors' cognitive bias, this paper provides a quantitative measurement of the bias based on the conditional estimates of diffusion processes.
  We prove the asymptotic properties of the estimates, which helps to interpret the investment behaviors that if a feedback trader finds a security perform better than his expectation, he will expect the future return to be higher, while in the long term, this bias will converge to zero.
  Furthermore, the observed deviations between the return forecast and its realized value lead to adaptive expectations in reality, for which we raise an exponential smoothing model as an adjustment method.
  In the empirical study on the stock market in China, we show the effectiveness of the ES method in bringing the biased expectation closer to the fundamental level,
  and suggest that the feedback traders, who are often over-optimistic about the return, are likely to suffer from downside risk and aggravate the speculative bubbles in the market.  
\end{abstract}


\section{Introduction}
Positive feedback trading refers to an investment behavior which buys when prices rise and sells when prices fall.
It can be the result of various trading strategies, such as portfolio choice based on extrapolative expectations, the use of stop-loss orders, purchases on margin which are liquidated when the stock drops below a certain point, as well as dynamic trading strategies such as portfolio insurance.

There is substantial literature suggesting the evidence of positive feedback trading.
The work of Andreassen and Kraus(1990)\cite{andreassen_judgmental_1990} provides an experimental evidence. There experiments showed that although the subjects didn't chase the trend when the stock price changed little, they universally switched to chase the trend when the prices exhibited a significant change, buying more when prices rise and selling when prices fall.
A similar survey evidence comes from the work of Frankel and Froot(1988)\cite{Frankel1988} who point out that typical forecaster expects the dollar to continue to appreciate over a period of a month or less simply because it has been appreciating in the recent past.
Koutmos and Saidi(2001)\cite{koutmos_positive_2001} test the presence of positive feedback trading in the stock index returns of six Asian emerging stock markets, including Hong Kong, Malaysia, Singapore, and make an conclusion that in all six markets it is found that feedback back trading is an important factor in determining short-term movements in stock returns.
Similarly, Antoniou et al(2005)\cite{antoniou_index_2005} imply that positive feedback trading is observed across six major stock markets France, Germany, Japan, the UK, and the US.
Daníelsson and Love(2006)\cite{danielsson_feedback_2006} prove the existence of feedback trading at high frequencies.

Academic literature distinguishes between two types of speculators: rational speculators who trade on fundamentals and are likely to dampen excessive price fluctuation, and noise traders.
Although sometimes positive feedback traders are rational investors who pursue strategies like the dynamic hedging, for most of the time they are regarded as noise traders 
who, due to incomplete rationality, can be destabilizing by buying those overpriced and sell those underpriced, and thus moving prices away from the rational fundamentals.
Literature has studied comprehensively on this destabilizing power. While most arguments say essentially that rational speculators will "buck the trend" and by doing so bring prices closer to the fundamental values,
De Long et al(1990)\cite{de_long_positive_1990}, on the contrary, argue that if feedback trading exists, then rational speculators can further destabilize the market. 
This is because rational speculators who expect some future buying by noise traders would buy today in the hope of selling at a higher price tomorrow, which can further stimulate positive feedback traders and so move prices even further away from fundamental values than they would go in the absence of rational speculators. 
In addition, there are researches focusing on stabilizing power of the derivatives markets against those feedback traders. Antoniou et al(2005)\cite{antoniou_index_2005} show evidence with the view that the existence of futures markets reduces the impact of positive feedback trading and thus helps to stabilize prices. 
However, empirical research has so far reached no firm conclusion about the stabilizing power of other markets.

Another field of interest the literature has focused on is the relationship between positive feedback trading and investor sentiment.
Investor sentiment is a belief or an expectation about the future discounted cash flow and risk that is not justified by the information at hand.
Brown and Cliff(2005)\cite{brown_investor_2005} find that irrational sentiments of investors do affect asset price levels, resulting in asset valuation errors.
Dai and Yang(2018)\cite{dai_positive_2018} are among the first few who directly relate investor sentiment to feedback trading. 
They argue that positive feedback traders are more likely to trade when their sentiment is relatively high or low and tend to select the moment when most securities move together to erase the risk of choosing one security from many.
They also reveal a new pattern of feedback traders' behavior that feedback traders do not trade on every fluctuation of security price. Although feedback traders are a kind of noise trader who lacks complete rationality, they use a “rule of thumb” to trade.

While the concept of investor sentiment generally discusses investors' optimism or pessimism about security return, cognitive bias, as a common form of investors sentiment, specifically focuses on investors' certain way of thinking and acting and the consequent systematic deviations from a standard of rationality or good judgment.
However, little research has focused on the feedback trading and its underlying cognitive bias. 
In particular, these are not any quantitative methods to precisely provide a theoretical measurement of the consequence of cognitive bias and positive feedback trading.
This paper specifically focuses on a statistical method to evaluate the irrationality of positive feedback trading.
The basic intuition is that we attribute feedback trading behavior to cognitive bias, assuming that uninformed and irrational investors tend to base their investment decision-making procedure irrationally on the stock past performance.
In other words, a well-performed security leads to over-optimism in its future return and thus attracting investment, while a bad performance results in investors' aversion.
The model we used is another innovation in this field.
Existed literature commonly use time series model VAR or its modifications or regression model to fit the return, and use GARCH to model the volatility (see Baker and Stein(2004)\cite{baker_market_2004}, Brown and Cliff(2005)\cite{brown_investor_2005}, Kurov(2008)\cite{kurov_investor_2008}, Dai and Yand(2018)\cite{dai_positive_2018}).
However, the theory of stochastic process provides a new method to model the assets return and has shown great fitness.
The following sections will show that, expressing investors' cognitive bias in the form of conditional estimates of diffusion process, we are able to explain and simulate the magnitude of cognitive bias and positive feedback trading.
Furthermore, we propose an exponential smoothing method to adjust the biased expectation, bringing the return forecast closer to the rational fundamental level.

The remainder of this paper is organized as follows. Section 2 introduces the basic framework of the assets pricing stochastic model based on diffusion processes.
Section 3 discusses the parameter estimates of the diffusion process, proposes the mathematical expectation conditional on the security's past return performance, and further discusses the asymptotic properties together with financial interpretations.
Section 4 raises a time series method to adjust the biased return expectation.
Finally, section 5 presents the results of an empirical study in Chinese stock market, showing the effectiveness of the proposed adjustment method. Section 6 concludes the paper. 

\section{Modeling Framework}

In this section, we introduce the modeling framework and define notations in this paper.

Denote the assets price as a diffusion process $\{A_t,t\ge 0\}$ given by
$$
    A_t=A_0+\int\nolimits_{0}^{t} \mu X_sds+\int\nolimits_0^t\sigma X_sdW_s.
$$
Its more commonly used stochastic differential equation form is written as
\begin{equation}
    dA_t=\mu A_tdt+\sigma A_tdW_t,
    \label{stoeq}
\end{equation}
where $\mu$ and $\sigma$ are deterministic parameters satisfying $\sigma >0$. $\mu$ is the drift coefficient, representing
the average return of the price $A_t$ over $[0,t]$. $\sigma$ is the diffusion coefficient,
representing the volatility of the price, and its value usually ranges from $0.1$ to $0.6$.

Given an initial value $A_0=a_0$, stochastic differential equation (\ref{stoeq}) has the only strong solution
\begin{equation}
    A_t=A_0 \exp\left(\left(\mu-\frac{\sigma^2}{2}\right)t + \sigma W_t\right). 
    \label{solutionA}
\end{equation}
That is, the assets price $A_t$ is a geometric Brownian motion with drift coefficient $\mu$ and diffusion coefficient $\sigma$.

Logarithmic transformation of $A_t$ simplifies the solution (\ref{solutionA}). Denote the logarithmic process as $\{Z_t, t>0\}$, given by
\begin{equation}
    Z_t=ln(A_t)-ln(A_0)=\left(\mu - \frac{\sigma^2}{2}\right)t+\sigma W_t.
\end{equation}
Then $Z_t$ is a Brownian motion with drift coefficient $\mu -\frac{\sigma^2}{2}\triangleq\nu$ and diffusion coefficient $\sigma$, having properties
(1)$Z_0=0$, and (2) $Z_t-Z_0\sim N(\nu t, \sigma^2 t)$ where $N(\cdot)$ represents Gaussian distribution.
For notation simplicity, we focus on $\nu$ instead of $\mu$ in the following discussion in this paper, while the computation results of $\mu$ can be derived by adding a constant $\sigma^2 / 2$ to the corresponding results of $\nu$.

Let $R_i$ be the return over $[t_i,t_{i-1}]$, defined as
$$
    R_i=Z_{t_i}-Z_{t_{i-1}}=\nu \left(t_i-t_{i-1}\right)+\sigma \left(W_{t_i}-W_{t_{i-1}}\right),\quad i=1,2,3,.... 
$$
It can be shown that $R_i$ satisfies (1) $R_i \sim N\left(\nu \left(t_i-t_{i-1}\right),\sigma^2 \left(t_i-t_{i-1}\right)\right)$, and (2) $R_i \bot R_j,\forall i\neq j$.

Write $\widetilde{R}_T$ as the total return over the whole time period $[0,T]$:
$$
    \widetilde{R}_T=Z_T-Z_0.\footnote{Although a typically easier representation is $\widetilde{R}_T=Z_T$, to ensure the application of the model in the cases where $Z_0\neq 0$, we still use the general form in our discussion.}
$$

\section{Estimates of the Diffusion Process}
In this section, we introduce the estimates of the parameters of the diffusion process,
and then propose a conditional expectation method which measures investor's cognitive bias on return expectation.

\subsection{Unconditional Estimates of Diffusion Process}
We focus on the time period $[0,T]$ constructed by $n$ discrete time intervals $[t_i.t_{i-1}],i=1,2,...,n$ which, for simplicity, have equal length $h:=T/n=t_i-t_{i-1}$. Then $R_i\sim N(\nu h, \sigma^2 h)$.

Based on the $n$ observations $r_1,r_2,...,r_n$ of $R_{t_1},R_{t_2},...,R_{t_n}$ at time points $0<t_1<t_2<...<t_n=T$, the estimates of $\nu$ and $\sigma^2$ are given by
$$
    \hat{\nu}=\frac{1}{nh}\sum_{i=1}^n r_i=\frac{1}{T}(z_T-z_0),\quad \hat{\sigma}^2=\frac{1}{(n-1)h}\sum_{i=1}^n(r_i-\bar{r})^2,
$$
respectively, where $z_{t_i}\;(i=0,1,...,n)$ denote the observations of the assets logarithmic prices $Z_t$, and $\bar{r}=\sum_{i=1}^n r_i/n$.

\subsection{Parameter Estimates Given Specific Return Performance}

\subsubsection{Conditional Expectation of the Drift Coefficient}

\begin{proposition}
    Given that the return $\widetilde{R}_T$ on $[0,T]$ is higher than a constant $C$, the conditional expectation of $\nu$ is 
    $$
    E[\hat{\nu}\,|\,\widetilde{R}_T>C]=\frac{1}{P\{\widetilde{R}_T>C\}}(2\pi T)^{-\frac{1}{2}}\left[  \sigma exp\left(-\frac{(C-\nu T)^2}{2\sigma^2T}\right) +\frac{\nu}{\sigma} \int\nolimits_{C}^{\infty} exp\left(-\frac{(y-\nu T)^2}{2\sigma^2T}\right)dy \right].
    $$
    On the other hand, given that $\widetilde{R}_T$ is lower than a constant $C$, the conditional expectation of $\nu$ is 
    $$
    E[\hat{\nu}\,|\,\widetilde{R}_T\le C]=\frac{1}{P\{\widetilde{R}_T\le C\}}(2\pi T)^{-\frac{1}{2}}\left[ - \sigma exp\left(-\frac{(C-\nu T)^2}{2\sigma^2T}\right) +\frac{\nu}{\sigma} \int\nolimits_{-\infty}^{C} exp\left(-\frac{(y-\nu T)^2}{2\sigma^2T}\right)dy \right].
    $$
    Here
    $$
    P\{\widetilde{R}_T \le C\} =1-P\{\widetilde{R}_T>C\}=\Phi\left(\frac{C-\nu T}{\sigma \sqrt{T}}\right),
    $$
    and $\Phi(\cdot)$ represents the cumulative density function of standard Gaussian distribution. (Proof is shown in appendix \ref{appen1}.)
\end{proposition}\label{pro1}

Define 
\begin{equation}
    E[\hat{\nu}\,|\,\widetilde{R}_T>C] - \nu \label{defbias}
\end{equation}
(or $E[\hat{\nu}\,|\,\widetilde{R}_T \le C] - \nu$) as the bias, then this proposition provides a measurement of the bias of the parameter estimates in diffusion process pricing model due to the past return performance, evaluating the incomplete rationality of uninformed investors.

In reality, the constant $C$ can be the market average return or the expected return of an individual security, such as the one given by capital asset pricing model (CAPM), and $T$ is usually one year or a half year.
Investors commonly use the estimator $\hat{\nu}$ based on historical data to predict the assets price in the next time period, and accordingly make their investment decisions. 
Suppose $C$ is the market average return, then this proposition indicates the bias on return expectation resulting from the individual security's return being higher or lower than the market average level;
similarly, suppose $C$ is the expected return of an individual security, then it indicates the bias resulting from the the security's real return being higher or lower than the expected level.

\subsubsection{Simulations}

In this section, we visualize the relationships between the conditional expectation in proposition 1 and the parameters $\mu$ and $C$. 

\begin{figure}[htbp]
    \begin{minipage}[t]{0.5\linewidth}
    \centering
    \includegraphics[scale=0.3]{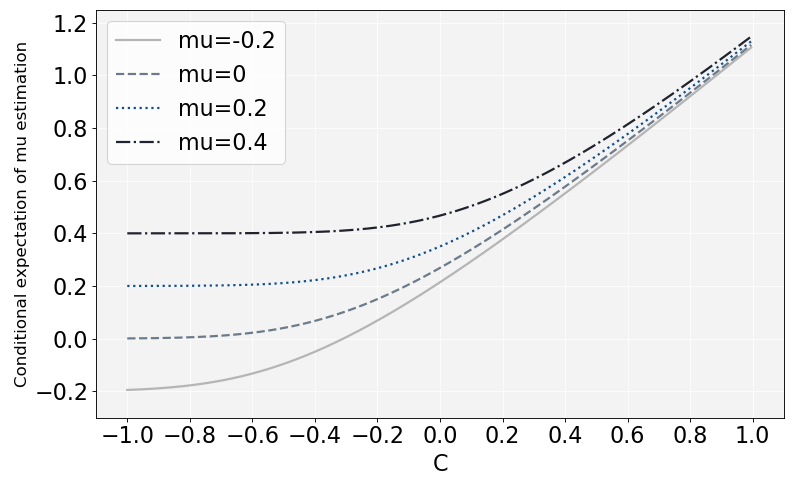}
    \end{minipage}%
    \hfill
    \begin{minipage}[t]{0.5\linewidth}
    \centering
    \includegraphics[scale=0.3]{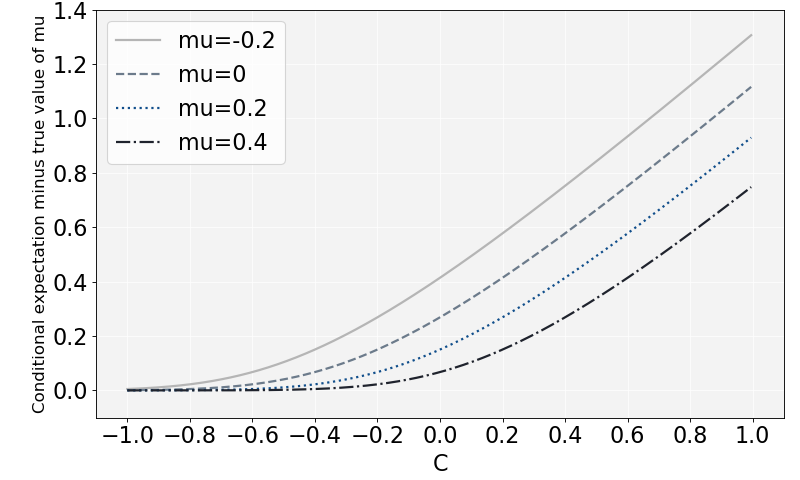}
    \end{minipage}
    \caption{Expectation of return given $\widetilde{R}_T>C$ (left), and the bias of the expectation given $\widetilde{R}_T>C$ (right).}
    \label{fig1}
\end{figure}

Figure \ref{fig1} show the variation in $E[\hat{\mu}\,|\,\widetilde{R}_T>C]\;(= E[\hat{\nu}\,|\,\widetilde{R}_T>C] + \sigma^2 / 2)$ and the corresponding bias defined in (\ref{defbias}), with $T=1,\sigma=0.3$. It shows that for a fixed $\mu$, a higher $C$ results in a higher $E[\hat{\mu}\,|\,\widetilde{R}_T>C]$. In other words, investors will expect a security to perform better in the future if it had a better return performance before, even though it is a fixed security with a certain fixed return rate. On the other hand, for a fixed $C$, a higher $\mu$ results in $E[\hat{\mu}\,|\,\widetilde{R}_T>C]$ being closer to its true value and accordingly a lower bias. It implies that the fact that a security's return is higher than the market average level or its own expected level won't provide investors much new information, if it is a security already having a high return rate, and thus won't encourage a significantly high expectation in the future. If $C=-1$, then $E[\hat{\mu}\,|\,\widetilde{R}_T>C] \equiv \mu; (\forall \mu)$, which is the trivial case where a security's return is higher than $-100\%$, providing no actual information.

\begin{figure}[htbp]
    \begin{minipage}[t]{0.5\linewidth}
    \centering
    \includegraphics[scale=0.3]{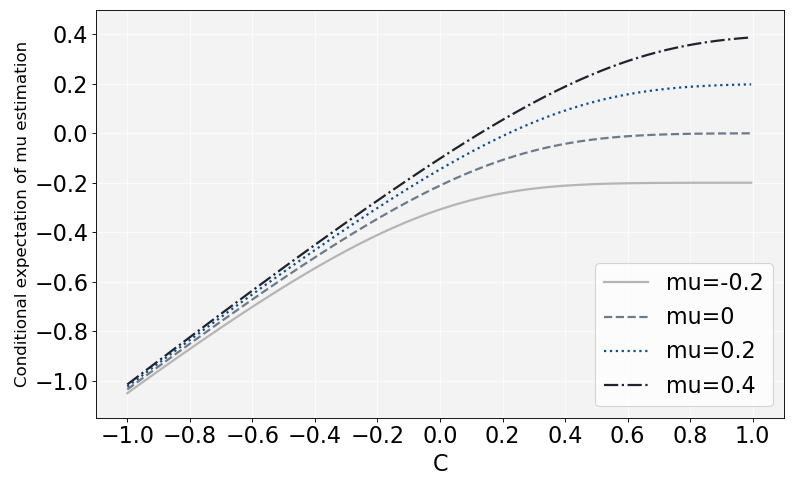}
    \end{minipage}%
    \hfill
    \begin{minipage}[t]{0.5\linewidth}
    \centering
    \includegraphics[scale=0.3]{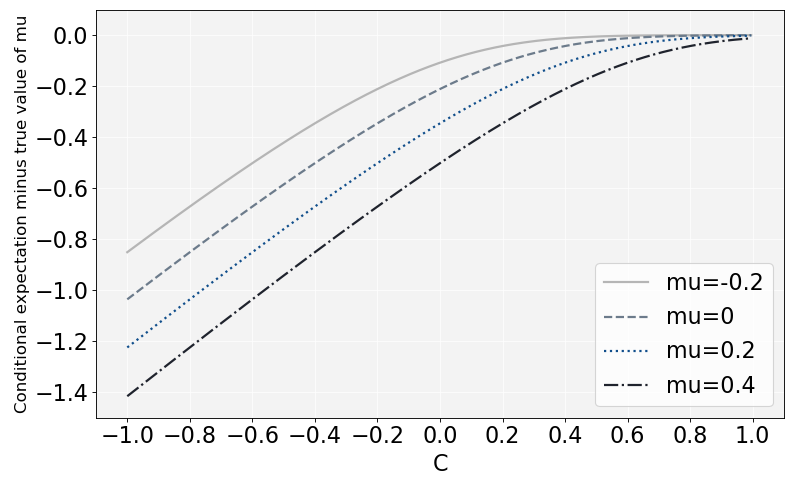}
    \end{minipage}
    \caption{Expectation of return given $\widetilde{R}_T \le C$ (left), and the bias of the expectation given $\widetilde{R}_T \le C$ (right).}
    \label{fig2}
\end{figure}

Similar interpretations can be done for the cases where $\widetilde{R}_T \le C$, as shown in figure \ref{fig2}. To avoid redundancy, we do not go into the details.

\subsubsection{Asymptotic Distribution of the Conditional Expectation with respect to time}

\begin{proposition}
  The limits of the conditional expectations with respect to the time $T$ are given by
  $$
  \lim_{T\to +\infty}E[\hat{\nu}\,|\,\widetilde{R}_T>C]=
  \begin{cases}
    \nu,&\text{if }\nu > 0,\\
    0,&\text{if }\nu \le 0.
  \end{cases}
  $$
  $$
  \lim_{T\to +\infty}E[\hat{\nu}\,|\,\widetilde{R}_T\le C]=
  \begin{cases}
    0,&\text{if }\nu \ge 0,\\
    \nu,&\text{if }\nu < 0.
  \end{cases}
  $$
  (Proof is shown in appendix \ref{appen2}.)
\end{proposition}

This proposition explains that the bias on return expectation can be eliminated if the time interval $[0,T]$ is wide enough.
Intuitively, the price of any security with $\nu>0$ (or $\mu > \sigma^2 / 2$) has a long-term growing tendency, then its return will necessarily grow higher than an arbitrary constant $C$.
Similarly, $\nu\le 0$ (or $\mu \ge \sigma^2/2$) represents a long-term decreasing tendency, leading to the security's return necessarily declining to a level lower than an arbitrary constant $C$.
Hence, the conditional expectation of return converges to the true value when the interval length $T$ goes to the infinity. 
Besides, the convergence depends on the volatility $\sigma$, which is reasonable in that a security with a strong volatility requires a high return rate to ensure a growing tendency in the long run.

In addition to the above stochastic analysis, the financial intuition is that a price movement in a given direction can lead to a trader to revalue the asset in the same direction, especially in a short period of time when they haven't yet attained sufficient information about the fundamental cause.
This proposition provides a statistical support for the view that the positive feedback traders are important for the short-run price behavior of an asset market while in the long run, they tend to revert to fundamentals. 
In fact, literature has shown consistent evidence from different perspective. 
For example, De Long et al(1990)\cite{de_long_positive_1990} discovered positive serial correlation at short horizons and negative serial correlation at longer horizons which may arise from the combination of speculators' anticipatory trades and positive feedback trading strategies.
Cohen and Shin(2002)\cite{cohen_positive_2002} suggest that in US Treasury market, trades and price movements appear likely to exhibit positive feedback at short horizons, particularly during periods of market stress.

\subsubsection{Conditional Expectation of the Diffusion Coefficient}

\begin{proposition}
    The estimator $\hat{\sigma}^2$ converges to the true value $\sigma$ as sample size $n$ increases. That is,
    $$
    \hat{\sigma}^2\to \sigma\;(n\to +\infty).
    $$
    (Proof is shown in appendix \ref{appen3}.)
\end{proposition}

The parameter's conditional estimates in diffusion process usually focuses on drift coefficient instead of diffusion coefficient (see Amaya et al(2019)\cite{amaya_maximum_2019}, Li et al(2004)\cite{li_conditional_2004}, Merton(1980)\cite{merton_estimating_1980}) due to the facts that
(1) the sum of square in $\hat{\sigma}^2$ makes computation too complex to solve,
and (2) the above proposition suggests that an increasing the sampling frequency provides a highly reliable estimate of $\sigma^2$, (i.e., $\hat{\sigma}^2$ is a consistent estimate of $\sigma$).
Additionally, an intuition also supports this argument. That is, the past return performance will influence investors' expectation on future return, being independent of the security's volatility.
Hence, this paper will not discuss the effect of past return performance on volatility expectation, and will use high-frequency daily price data to estimate $\sigma^2$ in the following empirical study.

\section{Adjustment of the Return Expectation}

The above discussion assumes that past return performance impacts the future return expectation, according to which trend-chasing investors will make their investment decision. 
However, the real return are always different from the expected one, 
and thus investors in reality will make adjustment to their current expectation based on the difference (called bias in this paper), between the deviated expectation made in the past and the real performance of the security at the corresponding time point.
In other words, we assume investors will adopt adaptive expectations in reality.
In this section, we raise a time series model to adjust investors expectation on return to bring it closer to the fundamental level.

\subsection{Exponential Smoothing Method}

Exponential smoothing is a time series forecasting method for univariate data.
Time series methods like the Box-Jenkins ARIMA family of methods develop models where the prediction is a weighted sum of past observations.
Exponential smoothing forecasting methods are similar in that a prediction is a weighted sum of past observations, but the model explicitly uses an exponentially decreasing weight for past observations.
Specifically, past observations are weighted with a geometrically decreasing ratio.

There are three main types of exponential smoothing time series forecasting methods: single exponential smoothing, double exponential smoothing, and triple exponential smoothing.
In the empirical study in section 5 we have implemented all of them and showed that single exponential smoothing performs well enough (though not written in this paper). Thus we only introduce this simplest version here. 

Single exponential smoothing is a time series forecasting method for univariate data without a trend or seasonality. Its formula is given by:
\begin{equation}
    F_{t+1}=\alpha Y_t+(1-\alpha)F_t, \label{expsm}
\end{equation}
where $F_t$ is the forecast at time $t$ and $Y_t$ is the observation at time $t$. The only parameter involved, $\alpha$, is called smoothing factor or smoothing coefficient,
which controls the rate at which the influence of the observations at prior time steps decays exponentially. 
An $\alpha$ value close to 1 indicates fast learning (that is, only the most recent values influence the forecasts), whereas a value close to 0 indicates slow learning (past observations have a large influence on forecasts).
Default values that have been shown to work well are in the range $0.1-0.2$ (see Shmueli and Lichtendahl(2016)\cite{shmueli2016practical}).

Equivalently, equation (\ref{expsm}) can be expressed as
$$
F_{t+1}=\alpha Y_t+(1-\alpha)\alpha Y_{t-1}+(1-\alpha)^{2}\alpha Y_{t-2}+\cdots +(1-\alpha)^{n}\alpha Y_{t-n}+\cdots +(1-\alpha)^{t}F_{1},
$$
which shows that an exponential smoother is a weighted average of all past observations with exponentially decaying weights. 

\section{An Empirical Study: The Biased Expectation on Return in China Stock Market and its Adjustment}

This section shows an empirical study in the stock market in China from 2009 to 2018 as a demonstration of the cognitive bias and its adjustment.

\subsection{Notations and Assumptions}

Recall in section 2 we introduce a constant $C$ denoting a benchmark for the return. Here we specifically let $C$ be the market equilibrium return rate in the capital asset pricing model (CAPM).
Then investors' forecast consists two parts: (1) expectation conditional on whether a stock performs better or worse than the theoretical market equilibrium level, biased due to positive feedback trading,
and (2) adjustments due to adaptive expectations, trying to bring the expectations closer to the fundamental level.

Here we define some necessary notations and assumptions as follows:
\begin{enumerate}
    \item Let $C$ be the market's average expectation on the return of an individual stock which is decided by CAPM. That is, $C=E(r)=r_f+\beta [E(r_M)-r_f]$.
    \item Define period $i$ to be the time interval $[T_i,T_{i+1}],i=0,1,2,...$. Denote the return on period $i$ by $\widetilde{R}_{T_i}$ and the unconditional estimates based on observations on period $i$ by $\hat{\nu}_i$.
          We assume that only when the return $\widetilde{R}_{T_i}$ of a stock on period $i$ is higher than $C$ will it attract investors on period $i+1$, otherwise investors won't invest in this stock. 
          Therefore, we define investors' expected return at period $i+1$ as $\widetilde{\nu}_{i+1} := I_{\{\widetilde{R}_{T_i}>C\}}\cdot E[\hat{\nu}_i\,|\,\widetilde{R}_{T_i}>C]$, where$I_{\{\widetilde{R}_{T_i}>C\}}$ is the indicator function.
    \item Define the bias as $\widetilde{\nu}_{i}-\hat{\nu}_i$\footnote{In section 2 we define the bias be $E[\hat{\nu}\,|\,\widetilde{R}_T>C] - \nu$. In fact the definition of bias as $\widetilde{\nu}_{i}-\hat{\nu}_i$ is essentially consistent to the definition in section 2. Here in the empirical study, 
              the true value of the population parameter $\nu$ is unknown, thus we substitute it with the sample estimator $\hat{\nu}_i$. And $\widetilde{\nu}_{i}$ is exactly the same as $E[\hat{\nu}\,|\,\widetilde{R}_T>C]$ except for an additional indicator function.}
          if an investment is made on period $i$, while if no investment is made, then let bias to be $0$. Then the following empirical study will use the sum of the squared deviation given by $\sum_{i=1}^{N} \left(\widetilde{\nu}_{i} - \hat{\nu}_{i}\right)^2$ where $N$ is the total time length, to evaluate the bias and its adjustment.
    \item We discuss the bias of return expectation and its adjustment on a basis of year, with the total time length being 10 years. That is, in unit of years, $T_i-T_{i-1}=1$ and $N=10$. 
\end{enumerate}

\subsection{Conditional Expectation on Return}

We randomly selected 10 stocks from CSI 300 Index which have complete trading data in the past ten years from 2009 to 2018 and cover the samples from both Shanghai and Shenzhen markets in China.
Then the necessary parameters in CAPM were collected, including the market risk free rates and the beta of each individual stock. 
We constructed the deviation series, conducted the relevant computation defined as above,
and table \ref{table1} shows stock 600519 as an example of the computation results.

\begin{table}[htbp]
	\centering
	\caption{Computation of different return estimates}
	\begin{threeparttable}
	\begin{tabular}{crrcrr}
		\toprule[0.8pt]
        No    & \multicolumn{1}{c}{$\hat{\nu}$} & \multicolumn{1}{c}{$\widetilde{\nu}$} & No    & \multicolumn{1}{c}{$\hat{\nu}$} & \multicolumn{1}{c}{$\widetilde{\nu}$} \\
		\midrule
		1     & 0.4469 & 0.0000 & 6     & 0.1226 & 0.2063 \\
		2     & 0.3329 & 0.4207 & 7     & 0.4085 & 0.4304 \\
		3     & -0.1529 & 0.0000 & 8     & 0.1081 & 0.2038 \\
		4     & 0.0513 & 0.1401 & 9     & 0.0073 & 0.0829 \\
		5     & 0.0883 & 0.1703 & 10    & 0.1358 & 0.1646 \\
		\bottomrule[0.8pt]
	\end{tabular}%
  \begin{tablenotes}
	  \footnotesize
	  \item Stock Code: 600519
  \end{tablenotes}
  \end{threeparttable}
	\label{table1}%
\end{table}%

\subsection{Adjustment of Return Expectation}

A simplest adjustment method is that, when making forecast on the next period, we simply add or subtract the deviation between the forecast and the observation on the last period.
We implemented this method, and, taking stock 002032 for example, the sum of the deviation without and with adjustment are 0.2266 and 0.1840 respectively. 
Therefore, employing this simple method can effectively make the forecast more reliable.

However, only based on deviation on the last period, the adjustment does not consider much past information, while in reality, investors will take account of all of the information in the past to adjust their forecast now. That is where the ES method we proposed in section 4 comes into play.

\begin{figure}[htbp]	
	\centering
    \includegraphics[scale=0.5]{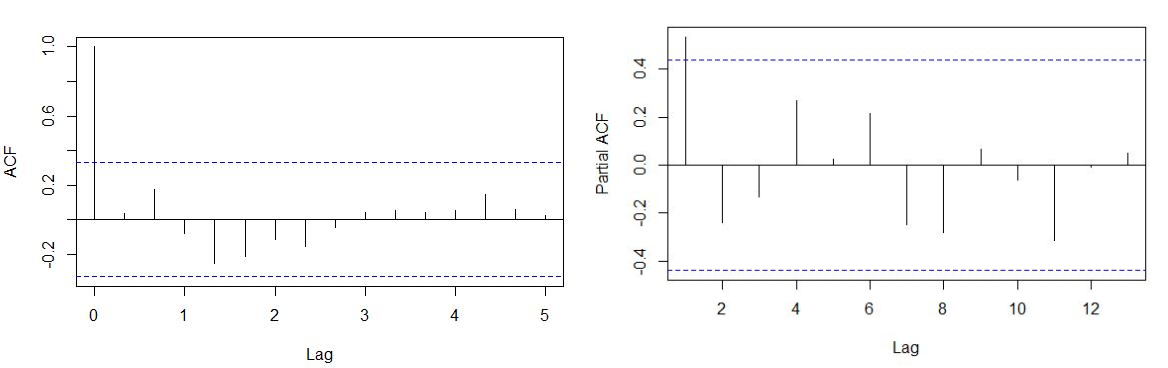}
	\caption{The autocorrelation plot and partial autocorrelation plot of stock:002032}
	\label{acf}
\end{figure}

Before modeling, we conducted tests of stationarity and Ljung-Box Q-test for each of the 10 stocks. 
Stationarity is required as it means that the statistical properties of a time series do not change over time, and many useful analytical tools and models, including the ES method, rely on it.
We do not go into the mathematical details here, but simply use ACF and PACF plots to test for stationarity. 
For example, the plots of stock 002032, as shown in figure \ref{acf}, proves the stationarity of this time series. 
The Ljung Box Q-test is a way to test for the absence of serial autocorrelation, up to a specified lag $k$. The null hypothesis of the test, is that the autocorrelations for the chosen lags are all zeros; or in simpler words, the time series being tested is white noise, containing no meaningful information.
We proved that the p-values of Ljung Box Q-test are all under 0.05, meaning that with significance level 0.05, the 10 time series are all worth modeling.

\begin{figure}[htbp]	
	\centering
    \includegraphics[scale=0.45]{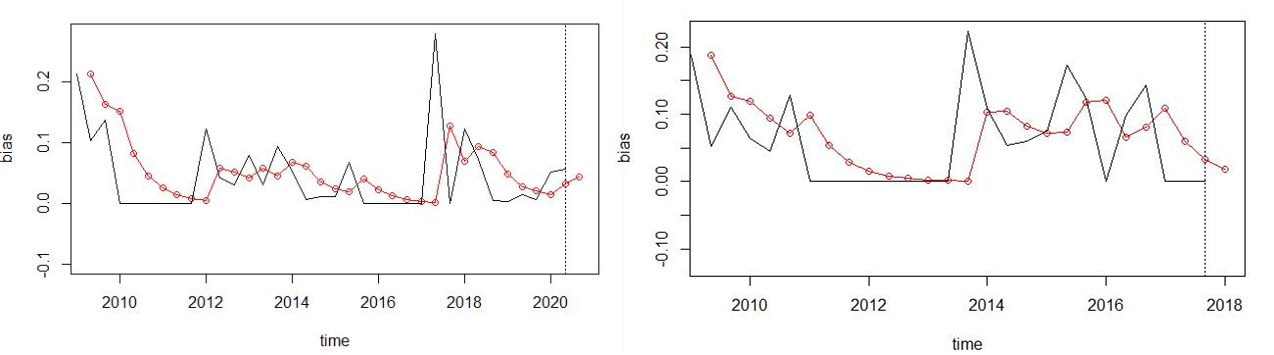}
	\caption{Adjusted expectations based on exponential smoothing of stock 600196 (left) and 600061 (right), where the black curves show the deviation series and the red curves show the adjusted expectations. }
	\label{stock}
\end{figure}

After the stationarity tests and Ljung-Box Q-test, we fitted the time series with exponential smoothing model.
Taking stock 600196 and 600061 as examples, figure \ref{stock} shows the deviation series (in black) from 2009 to 2018 and its smoothing results (in red).
We can see that ES method provides a smoothed curve, denoting a more rational return forecast.

Finally, we tested the forecast accuracy of the ES method. We made forecast on return in January 2019 which was not covered in the sample period.
The results are shown in table \ref{table3}. The second column shows the parameter estimates from historical data in January 2019 and the third column shows the conditional expectations.
The fourth and the fifth columns compare the two adjustment methods, i.e., the simple adjustment and the ES adjustment, respectively. 
The last three columns calculate the squared deviation between the historical data and the adjusted forecast using the 3 different forecast methods in columns 3 to 5.
To make the results clearer, we calculated the sum of the squared deviation and showed the results in the final row.
As a result, the simple adjustment method alleviates the forecast deviation partially, while the ES method significantly lowers the deviation by 51.78\%, showing high effectiveness.

\begin{table}[htbp]
	\centering
	\caption{Return forecast and adjustment results of the 10 stocks}
	\begin{tabular}{cccccccc}
		\toprule[0.8pt]
		\textbf{Stock code} & \textbf{$\hat{\nu}$} & \textbf{$\widetilde{\nu}$} & \textbf{SA} & \multicolumn{1}{c}{\textbf{ESA}} & \textbf{SD for $\widetilde{\nu}$} & \textbf{SD for SA} & \textbf{SD for ESA} \\
		\midrule[0.5pt]
		\textbf{002032} & 0.3288 & 0.3507 & 0.2965 & \multicolumn{1}{c}{0.2965} & 0.0005 & 0.0010 & 0.0010 \\
		\textbf{600028} & 0.1151 & 0.1626 & 0.1560 & \multicolumn{1}{c}{0.1165} & 0.0023 & 0.0017 & 0.0000 \\
		\textbf{600030} & -0.0593 & 0.0965 & 0.0769 & \multicolumn{1}{c}{0.0491} & 0.0243 & 0.0186 & 0.0117 \\
		\textbf{600061} & 0.0573 & 0.3192 & 0.3192 & \multicolumn{1}{c}{0.3016} & 0.0686 & 0.0686 & 0.0597 \\
		\textbf{600115} & -0.0344 & 0.1240 & 0.1240 & \multicolumn{1}{c}{0.1097} & 0.0251 & 0.0251 & 0.0208 \\
		\textbf{600352} & -0.0972 & 0.0709 & 0.0684 & \multicolumn{1}{c}{0.0039} & 0.0282 & 0.0274 & 0.0102 \\
		\textbf{600519} & 0.2818 & 0.2980 & 0.2979 & \multicolumn{1}{c}{0.2859} & 0.0003 & 0.0003 & 0.0000 \\
		\textbf{601727} & -0.1617 & 0.1443 & 0.0684 & \multicolumn{1}{c}{-0.1251} & 0.0936 & 0.0529 & 0.0013 \\
		\textbf{601898} & 0.1865 & 0.2723 & 0.2456 & \multicolumn{1}{c}{0.1088} & 0.0074 & 0.0035 & 0.0061 \\
		\textbf{600196} & -0.1331 & 0.0984 & 0.0952 & \multicolumn{1}{c}{0.0559} & 0.0536 & 0.0521 & 0.0357 \\
		\midrule
		\textbf{TOTAL} &       &       &       &       & \textbf{0.3038} & \textbf{0.2512} & \textbf{0.1465} \\
		\bottomrule[0.8pt]
	\end{tabular}%
	\label{table3}%
\end{table}%

\subsection{Discussions on Downside Risks and Speculative Bubbles}

Using conditional model as an measurement of investors' incomplete rationality and cognitive bias, this paper tries to illustrate the consequence of positive feedback trading in financial market.
Based on the empirical study, we can conclude that the positive feedback trading will lead to a positive bias on return expectation, mainly because those over-optimistic feedback traders ignore the downside risk.

Immature investors in emerging market such as China lack rational investment strategy, and tend to long after going up and short if dropping down. 
In particular, they often expect a well-performed stock to perform even better in the future in the hopes that they will also end up with positive outcomes. 
Although sometimes this optimism results in a positive outcome, more often it will lead to the adverse cases.
Due to various reasons such as the lack of enough information and incomplete irrationality, investors ignore the underlying causes about why a stock could perform better than expectation, and thus are likely to suffer from downside risk.

Besides the risk that the positive feedback trading brings to those overconfident investors, it also encourages the speculative bubbles if more and more investors are attracted by the profit, breaking the market equilibrium and efficiency.
A good example is the stock market turbulence in China in 2015-2016, where the rocketing stock index made investors more and more excited, even including those well-trained institutional investors.
As discussed in the work of Brown and Cliff(2005)\cite{brown_investor_2005}, market regulators and government officials should be concerned about the potential for market bubbles or “irrational exuberance” if a sudden change in sentiment translates into a negative wealth shock that depresses economic activity.

Although there is no firm conclusion in literature about whether adaptive expectations are rational or quite the contrary, such strategy does generally exist in real life and sometimes is effective in bringing the return forecast back to the fundamental level.
This paper proposes an exponential smoothing method and proves its efficiency, hoping investors to fully consider all of the information in the past and formulate their strategies on a strong theoretical basis.

\section{Conclusion}

This paper focuses on the positive feedback trading and relates it to the underlying cognitive bias. We model the assets price with diffusion processes, and take into account the cognitive bias by computing the conditional estimates of the diffusion process given the past return performance. The discussion on the estimates asymptotic properties provides an interpretation of the investment behavior. 
It is shown that in the short term, a security's better performance than expected encourages a feedback trader to expect significantly higher on its return in the future; while in the long term, the bias converges back to zero. The magnitude of the bias depends on the security's real return level.

Secondly, considering adaptive expectations in reality, this paper introduces an exponential smoothing method to adjust the return expectation closer to its fundamental level.
Finally, we conduct an empirical study on 10 stocks from CSI 300 in China in 10 years from 2009 to 2018. 
We compare a simple adjustment method to the ES method in the empirical study and show the effectiveness of the ES method which can significantly lower the forecast deviation.
In conclusion, adopting positive feedback trading strategy, investors are often over-optimistic about the future return and may thus suffer from the downside risk. 
In addition, this irrationality plays a role in the speculative bubbles in the market.
Although cognitive bias is sometimes unavoidable, especially for individual investors who lack comprehensive investigations into the market, we suggest that investors adopt trading strategies with a reasonable theoretical support such as the ES method, and fully consider all the information in the past to guide their investment decisions.

\bibliographystyle{unsrt}  
\bibliography{references}

\section*{Appendix}
\appendix
\section{Proof of proposition 1}\label{appen1}

We provide the proof for $E[\hat{\nu}\,|\,\widetilde{R}_T>C]$, and the one for $E[\hat{\nu}\,|\,\widetilde{R}_T\le C]$ are similar.
\begin{proof}
    \begin{equation} \nonumber
      \begin{aligned}
        E[\hat{\nu}\,|\,\widetilde{R}_T>C]&=E\left[\frac{1}{T}(Z_T-Z_0)\,|\,Z_T-Z_0>C \right] \\
        & \xlongequal  [Y_T \sim N(\nu T, \, \sigma^2T)]{Y_T:=Z_T-Z_0} E\left[\frac{1}{T} Y_T \,|\, Y_T >C\right] = \frac{1}{P\{Y_T>C\}} \int\nolimits_{\{Y_T>C\}}\frac{1}{T}Y_TdP,
      \end{aligned}
    \end{equation}
    where
    \begin{equation} \nonumber
    P\{Y_T>C\} = 1-\Phi\left(\frac{C-\nu T}{\sigma \sqrt{T}}\right) = P\{\widetilde{R}_T>C \}
    \end{equation}
    since $Y_T \sim N(\nu T, \, \sigma^2T)$, and
    \begin{equation} \nonumber
      \begin{aligned}
        \int\nolimits_{\{Y_T>C\}}\frac{1}{T}Y_TdP & = \int\nolimits_{C}^{\infty} \frac{1}{T}y \cdot \frac{1}{\sigma \sqrt{T}} \phi \left(\frac{y-\nu T}{\sigma\sqrt{T}}\right)dy \\
        & = \frac{1}{\sigma T\sqrt{T}}(2\pi)^{-\frac{1}{2}}\int\nolimits_C^\infty y\cdot exp\left( -\frac{(y-\nu T)^2}{2\sigma^2T} \right) dy \\
        & = \frac{1}{\sigma T\sqrt{T}}(2\pi)^{-\frac{1}{2}} \left[ \sigma^2Texp\left(-\frac{(C-\nu T)^2}{2\sigma^2T}\right) + \nu T \int\nolimits_C^\infty exp\left(-\frac{(y-\nu T)^2}{2\sigma^2T}\right)dy \right] \\
        & = (2\pi T)^{-\frac{1}{2}} \left[ \sigma exp\left(-\frac{(C-\nu T)^2}{2\sigma^2T} \right) +\frac{\nu}{\sigma} \int\nolimits_C^\infty exp\left(-\frac{(y-\nu T)^2}{2\sigma^2T}\right)dy \right],
      \end{aligned}
    \end{equation}
    where $\Phi(\cdot)$ and $\phi(\cdot)$ represent the cumulative density function and probability density function for standard Gaussian distribution respectively.
\end{proof}
Besides, for clarifications, we make a substitution by letting $Y_T:=Z_T-Z_0$ for convenience, which, strictly speaking, is not necessary.
Moreover, we have $Z_0 \equiv 0$ by definition of $Z_T$ in this paper, and then $Y_T \equiv Z_T$. 
Therefore, in fact this substitution brings negligible improvement. 
However, we keep writing $Y_T=Z_T-Z_0$, allowing for different definitions in other literature. 
As long as the definition differs from the one in this paper simply in a constant term, $Y_T$ will follow a Gaussian distribution.

\section{Proof of proposition 2}\label{appen2}

We prove the convergence property of $E[\hat{\nu}\,|\,\widetilde{R}_T>C]$ with respect to $T$, and the one of $E[\hat{\nu}\,|\,\widetilde{R}_T\le C]$ are similar.
\begin{proof}
    \begin{equation} \nonumber
    \begin{aligned}
    E[\hat{\nu}\,|\,\widetilde{R}_T>C] & =\frac{1}{P\{\widetilde{R}_T>C\}}(2\pi T)^{-\frac{1}{2}}\left[  \sigma exp\left(-\frac{(C-\nu T)^2}{2\sigma^2T}\right) +\frac{\nu}{\sigma} \int\nolimits_{C}^{\infty} exp\left(-\frac{(y-\nu T)^2}{2\sigma^2T}\right)dy \right] \\
    & = \frac{1}{1 - \Phi\left(\frac{C-\nu T}{\sigma \sqrt{T}}\right)}\cdot \left[ \frac{\sigma}{\sqrt{T}} \phi\left(\frac{C-\nu T}{\sigma \sqrt{T}}\right) +\nu \left(1-\Phi\left(\frac{C-\nu T}{\sigma \sqrt{T}}\right) \right) \right] \\
    & = \frac{1}{1 - \Phi\left(\frac{C-\nu T}{\sigma \sqrt{T}}\right)} \cdot \frac{\sigma}{\sqrt{T}} \cdot \phi\left(\frac{C-\nu T}{\sigma \sqrt{T}}\right) + \nu
    \end{aligned}
    \end{equation}
    Let $T\rightarrow +\infty$,
    \begin{enumerate}
    \item 
    If $\nu > 0$, then $\frac{C-\nu T}{\sigma \sqrt{T}}\rightarrow -\infty$. Then we have $\Phi\left(\frac{C-\nu T}{\sigma \sqrt{T}}\right) \rightarrow 0$ (by continuity of cdf $\Phi(\cdot)$) and $\frac{\sigma}{\sqrt{T}} \cdot\phi\left(\frac{C-\nu T}{\sigma \sqrt{T}}\right) \rightarrow 0$. Thus $E[\hat{\nu}\,|\,\widetilde{R}_T>C]\rightarrow \nu$.
    \item 
    If $\nu < 0$, then $\frac{C-\nu T}{\sigma \sqrt{T}}\rightarrow +\infty$. Similarly we have $1-\Phi\left(\frac{C-\nu T}{\sigma \sqrt{T}}\right) \rightarrow 0$ and $\frac{\sigma}{\sqrt{T}} \cdot\phi\left(\frac{C-\nu T}{\sigma \sqrt{T}}\right) \rightarrow 0$, and it can be easily showed that
    \begin{equation} \nonumber
    \lim_{T\to +\infty} \frac{\frac{\sigma}{\sqrt{T}} \cdot\phi\left(\frac{C-\nu T}{\sigma \sqrt{T}}\right)}{1-\Phi\left(\frac{C-\nu T}{\sigma \sqrt{T}}\right)} = \lim_{T\to +\infty} \frac{C^2 - \sigma^2 T - \nu^2 T^2}{CT + \nu T^2} = -\nu
    \end{equation}
    Thus, $E[\hat{\nu}\,|\,\widetilde{R}_T>C]\rightarrow 0$.
    \item 
    If $\nu = 0$, then $1 / (1 - \frac{C-\nu T}{\sigma \sqrt{T}})$ and $\phi\left(\frac{C-\nu T}{\sigma \sqrt{T}}\right)$ are bounded. Since $\frac{\sigma}{\sqrt{T}}\rightarrow 0$, we have $E[\hat{\nu}\,|\,\widetilde{R}_T>C]\rightarrow 0$.
    \end{enumerate}
\end{proof}

\section{Proof of proposition 3}\label{appen3}
\begin{proof}
    \begin{equation} \nonumber
    \hat{\sigma^2}  = \frac{1}{(n-1)h}\sum_{i=1}^{n}(r_i-\bar{r})^2 
    \end{equation}
    Since $\bar{r} = \sum_{i=1}^{n}r_i / n = \frac{1}{n}(z_T-z_0)$, $r_i=z_{t_i} - z_{t_{i-1}}$, then
    \begin{equation} \nonumber
    \begin{aligned}
    \hat{\sigma}^2&=\frac{1}{(n-1)h}\sum_{i=1}^n \left[ (z_{t_i} - z_{t_{i-1}}) - \frac{1}{n}(z_T-z_0) \right]^2 \\
    & =\frac{1}{(n-1)h} \sum_{i=1}^n \left[ \left(\nu h + \sigma(W_{t_i} - W_{t_{i-1}})\right) - (\nu h +\frac{1}{n}\sigma W_T) \right]^2 \\
    & = \frac{1}{(n-1)h} \sum_{i=1}^n  \left[ \sigma (W_{t_i} - W_{t_{i-1}}) - \frac{1}{n}\sigma W_T  \right]^2 \\
    & = \frac{1}{(n-1)h} \left[ \sigma^2 \sum_{i=1}^n (W_{t_i} - W_{t_{i-1}})^2 - \frac{2\sigma^2}{n}W_T(W_T-W_0) + \frac{\sigma^2}{n}W_T^2  \right] \\
    & = \frac{n\sigma^2}{(n-1)T} \sum_{i=1}^n (W_{t_i} - W_{t_{i-1}})^2 - \frac{\sigma^2}{(n-1)T}W_T^2 \\
    & \overset{L^2}{\rightarrow} \frac{\sigma^2}{T} \cdot T - 0 \quad (n\rightarrow +\infty)\\
    & = \sigma^2
    \end{aligned}
    \end{equation}
    where the limit exists due to the convergence of quadratic variation of Brownian motion in $L^2$.
\end{proof}

\end{document}